\documentclass[onecolumn,usenatbib,amsmath,amssymb,amsbsy]{mn2e}

\usepackage{natbib}
\def\X{{\mathrm{x}}}
\def\Y{{\mathrm{y}}}

\def\n{{\rm n}}
\def\p{{\rm p}}
\def\e{{\rm e}}
\def\s{{\rm s}}

\def\L{{\rm L}}

\def\be{\begin{equation}}
\def\ee{\end{equation}}
\def\bea{\begin{eqnarray}}
\def\eea{\end{eqnarray}}

\newcommand{\pd}[2]{\frac{\partial {#1}}{\partial {#2}}}

\begin{document}

\title{Mutual friction in superfluid neutron stars}

\author[N. Andersson et al]{N.~Andersson$^1$, T. Sidery$^1$ and G.L. Comer$^2$ \\
$^1$ School of Mathematics, University of Southampton, 
Southampton SO17 1BJ, United Kingdom\\
$^2$ Department of Physics \& Center for Fluids at All Scales, 
Saint Louis University, St.~Louis,
MO, 63156-0907, USA}

\maketitle

\begin{abstract}
We discuss  vortex-mediated mutual friction in the two-fluid model for 
superfluid neutron star cores. Our discussion is based on the general formalism developed 
by Carter and collaborators, which makes due distinction between 
transport velocity and momentum for each  fluid. This is essential for an implementation
of the so-called entrainment effect, whereby the flow of one fluid imparts momentum in the other and vice versa.
The mutual friction follows by balancing the Magnus force that acts on the quantised 
neutron vortices with a resistive force due to the scattering of electrons off of the 
magnetic field with which each vortex core is endowed. We derive the form of the macroscopic
mutual friction force which is relevant for a model based on smooth-averaging over 
a collection of vortices. We discuss the coefficients that enter the expression for this force, and
the timescale on which the two interpenetrating fluids in a neutron star core are coupled.
This discussion confirms 
that our new formulation accords well with previous work in this area.
\end{abstract}

\section{Introduction}

A superfluid rotates by forming a dense array of quantised vortices. In the case of mature neutron stars, 
which are expected to exhibit large scale superfluidity since their core temperatures
are orders of magnitude below the Fermi temperatures for both neutrons and protons, 
these vortices should play a key role in determining the 
rotational dynamics. In fact, the sudden spin-up associated with the observed radio pulsar glitches \citep{lyne}, 
and the relaxation that follows,
is commonly viewed as strong evidence of transfer of angular momentum between a superfluid component
and the charged component to which the star's magnetic field is locked and which, presumably, 
is linked to the pulsar emission mechanism. Hence, the notion that a neutron star 
acts as a kilometer-sized superfluid system is strongly supported, both theoretically and 
observationally.  

The purpose of this study is to model vortex-mediated dissipation in the 
two-fluid paradigm for superfluid neutron stars. 
The common view is that the 
most important dissipation mechanism in a superfluid neutron star core originates from 
electrons scattering off of the magnetic fields associated with the individual 
vortex cores \citep{als84,mendII}. Key to this idea is, as discussed by  \citet{sss} 
(see also \citet{sed1}), the fact that the entrainment effect 
induces a flow in the proton fluid around each neutron vortex. 
This, in turn, generates a local magnetic field of the order of $10^{14}$~G off of which the 
electrons scatter dissipatively \citep{als84}. 
The outcome is a coupling between the 
neutrons and the interpenetrating conglomerate of charged particles. This mechanism has been considered in two 
inportant scenarios. First,  \citet{as88} have argued that it leads to the core fluids coupling 
on a timescale of $400-10^4$ rotation periods. (For alternative pictures, where the coupling timescale 
is significantly different, see \citet{sed2,sed3,sed4}.) As this is far faster than the relaxation timescale
following a Vela pulsar glitch, one can argue that the glitches cannot be associated with 
the core but rely on the conditions in the crust. Second,  \citet{mendII} discussed the fact that the 
mutual friction is also an important damping agent for neutron star oscillations. This is of 
key importance for potential gravitational-wave driven mode-instabilities (see \citet{andrev} 
for a recent review). In fact, the current thinking is that the mutual friction supresses the 
instability in the star's f-mode entirely \citep{lm_f}. The effect on the unstable r-modes is not quite
as devastating \citep{lm_r}, but it could well be that the mutual friction sets the most stringent 
constraints on the instability window also in this case.  

In recent work we have applied the general formalism for superfluid neutron stars in Newtonian gravity
developed by  \citet{rp1} (see also the closely related work by  \citet{cc1,cc2,cc3})
to a set of problems relevant for astrophysical neutron stars.  
In particular, we have considered rotating stellar configurations where the two fluids are allowed 
to spin at different rates  \citep{rotpaper,GRrot,PNC}. We have also discussed
the nature of the inertial modes of oscillation (of which the r-modes form a sub-class) \citep{inertial}, and
investigated the possibility that a two-stream instability may operate in a superfluid neutron star \citep{prl,2stream}.  
These papers demonstrate clearly  i) that the superfluid oscillation problem is richer than tends to be assumed, and ii)
that the entrainment effect (whereby the flow of one fluid imparts momentum in the other) plays a 
key role in determining the nature of the various modes of oscillation and 
the extent to which the various fluids partake in the pulsation.

Our previous studies were based on the equations that follow after smooth-averaging over a 
collection of vortices. This ``macroscopic'' approach does
not provide insight into the actual vortex dynamics. In order to devise a model 
for the required mutual friction force we must make connection between the large scale dynamics that 
we have previously considered
and the ``mesoscopic'' level, which is sufficiently resolved that individual vortices can be distinguished yet 
sufficiently coarse that we do not have to worry about ``microscopic'' quantum effects (other than the quantisation of vorticity).
Our analysis is based on the well-established procedure for deducing the mutual friction force in 
the case of superfluid Helium \citep{hv,bk,book1,book2}, and proceeds in 
three main steps: First we 
discuss the nature of the quantised vorticity, making the 
appropriate distinction between transport velocities and momenta. This is 
important if one wants to correctly account for the entrainment. 
Next we derive an expression for the Magnus force, which 
describes how a bulk flow imparts a force on the vortices (analogous to the so-called
Joukowski lift in standard fluid mechanics); see \citet{sonin} for a useful review. 
Finally, we derive an expression
for the vortex-mediated mutual friction force, which couples the
superfluid neutrons to the conglomerate of charged components. 
Having derived this expression we discuss the relevant coefficients
and compare our final results to previous ones in the literature. 

\section{The superfluid equations of motion}

We  take as our starting point the two-fluid equations 
derived by, for instance,  \citet{rp1} (see also \citet{monster}). 
In this description the number density of each fluid obeys the
 continuity equation
\begin{equation}
\pd{n_{\X}}{t} + \nabla_{j} (n_{\X} v_{\X}^{j}) = 0 \ . 
\label{eq:Cont}\end{equation}
Here, we distinguish between the  ``consitutent index'' $\X$  which (in the present context) 
can be either $\n$ or $\p$, and the spatial index $j$. 
The index $\n$ represents the superfluid neutrons while $\p$ corresponds to a
conglomerate of all charged particles  
(protons and electrons), which are expected to flow together 
due to electromagnetic coupling \citep{mendI}. In the following, repeated constituent indices
($\X$ and $\Y$) never imply summation while spatial indices $i$, $j$ and $k$ 
satisfy the Einstein summation convention.
 In Eq.~(\ref{eq:Cont}), $n_\X$ is the number density and $v_\X^i$ 
is the transport velocity. That is, $n_\X^i = n_\X v_\X^i$ represents the 
true number density current for species $\X$.  

Each fluid satisfies an Euler-type equation, which ensures the 
conservation of total momentum. For constituent $\X$ this equation can be written
\begin{eqnarray}\label{eq:MomEqu}
\left(\pd{}{t} + {v}^{j}_{\X}\nabla_{j} \right)   
\left[ {v}_{i}^{\X} + \varepsilon_{\X} w_i^{\Y\X} \right] + \nabla_{i} (\Phi + 
\tilde{\mu}_{\X}) 
+ \varepsilon_{\X}  w_j^{\Y\X} \nabla_{i} v^{j}_{\X} = 0 \ .
\end{eqnarray}
Here we have defined the relative velocity
\be
w_i^{\Y\X} =  {v}_{i}^{\Y} - {v}_{i}^{\X} \ . 
\ee
Furthermore,
\be 
\tilde{\mu}_\X = {\mu_\X \over m_\X} = { 1 \over m_\X} { \partial E \over \partial n_\X} \ ,
\ee
where $E$ is the internal energy of the system, is the relevant chemical potential per unit 
mass.  The entrainment is included via the coefficients
\be
\varepsilon_\X = 2 \rho_\X \alpha \qquad  \mbox{ where } \qquad \alpha = { \partial E \over \partial w_{\n\p}^2} \ ,
\ee 
and
$\Phi$ represents
the gravitational potential.
For a detailed discussion of these equations, see \citet{rp1,monster}.

Using the fact that the momentum per particle 
(which is canonically conjugate to the 
number current $n_\X^i$) of each fluid is defined as
\be
p_i^\X =  m_\X[ {v}_{i}^{\X} + \varepsilon_{\X} w_i^{\Y\X}]  \ , 
\label{mom_def}\ee
and identifying the spatial derivatives in (\ref{eq:MomEqu}) as 
components of the Lie derivative $\pounds_{v_\X}$ associated with the 
velocity $v_\X^i$, i.e. using
\be
\pounds_{v_\X} W_i = v_\X^j \nabla_j W_i + W_j \nabla_i v_\X^j  
\ee
which holds for a general co-vector $W_i$, we can rewrite
(\ref{eq:MomEqu}) as
\be
\left(\pd{}{t} + \pounds_{v_\X}  \right)  p^\X_i +  m_\X \nabla_{i} \left( \Phi + 
\tilde{\mu}_{\X} - { 1 \over 2} v_\X^2 \right)  = 0 \ . 
\label{mom_2}\ee

Let us now make contact with microphysics. In the case of a neutron superfluid 
the momentum $p^\n_i$ is related to the gradient of the phase $\chi_\n$ of the 
condensate wavefunction (the ``order parameter'') via
\be
p_i^\n = { \hbar \over 2 } \nabla_i \chi_\n \ ,
\ee 
where the factor of 2 is introduced since we are dealing with 
neutron Cooper pairs. Taking the curl of this relation, we see that the superfluid is 
generally irrotational. It follows from (\ref{mom_2}) that 
this property is conserved by the flow, see \citet{rp1}, and hence it is  natural that 
vortices are associated with quantised momentum circulation. To some extent, this contrasts 
with the  ``orthodox'' Landau 
formulation of superfluids (see the work of, for example, \citet{mendI,mendII}). 
In that paradigm $p_i^\n/m_\n$ is refered to as
the ``superfluid velocity'', $V^\s_i$ (say). 
Conceptually, this is somewhat confused
but there is no real risk of making significant mistakes as long as 
one does not try to account for entrainment. After all, 
if $\varepsilon_\n=0$ we trivially have $V^\s_i = v^\n_i$. As we will see, 
the situation when entrainment is considered is far from trivial and one must take some care 
in order to avoid inconsistencies.

If we introduce vortices in the superfluid
then it is easy to show that the circulation in the neutron 
fluid must be quantised. Integrating along a contour which encloses a single vortex
we have
\be
{\cal C} = \oint p^\n_i dx^i = \int (\epsilon^{ijk} \nabla_j p^\n_k)  dS_i = { h \over 2}  \ .
\label{n_circ}\ee
If  for simplicity we assume that the vortex is straight, and introduce
cylindrical coordinates ($r,\theta,z$) centered on the vortex, the neutron momentum 
can be represented by 
\be
{p}^\n_\theta =  \frac{{\cal C} }{2 \pi} \ .
\label{n_mom}\ee
It should be noted that, in an orthonormal basis, this corresponds to the more familiar
looking vortex solution $\vec{p} = {\cal C} \hat{e}_\theta/2\pi r$.
Formally, the vorticity of the flow is associated with the singularity at $r=0$. 
However, on a macroscopic scale we can meaningfully introduce a smooth averaged 
``rotation velocity'' by comparing the standard result
\be
2 \Omega_\X^i = \epsilon^{ijk} \nabla_j v_k^\X \ ,
\ee
to (\ref{mom_def}). Hence, on the macroscopic level we identify (assuming that the lengthscale considered is
sufficiently small that we can treat $\varepsilon_\n$ as a constant)
\be
\epsilon^{ijk} \nabla_j p^\n_k = 2 m_\n [ \Omega_\n^i + \varepsilon_\n (\Omega_\p^i - \Omega_\n^i)] \ .
\ee
As a result, we find that by enclosing $\cal N$ vortices we get 
\be
2 \pi \int_0^r m_\n [ \Omega_\n + \varepsilon_\n (\Omega_\p - \Omega_\n)] r dr = { {\cal N} h \over 2} \ ,
\ee
where we have used $\Omega_\X^i = \Omega_\X \hat{e}_z^i$. Defining the surface vortex density $n_v$ as
\be
n_v = { d {\cal N} \over 2 \pi r dr} \ ,
\ee 
we see that
\be
n_v \kappa = 2  [ \Omega_\n + \varepsilon_\n (\Omega_\p - \Omega_\n)] \ .
\label{n_v}\ee
It should be noted that this result, which clearly displays the interpenetrating nature of the two fluids,  
differs from the ``standard'' result (see for example \citet{als84}). Because of the
entrainment  effect, the number density of neutron vortices depends explicitly on the rotation of the 
proton fluid (this point was recently discussed also by  \citet{cc05}).

Let us now consider the force acting on a single neutron vortex. This means that the 
neutron momentum is represented by (\ref{n_mom}), since we focus on the mesoscopic scale. 
Meanwhile, there should (on this scale) be
no circulation in the proton momentum. We will motivate this assumption later when we account for the fact that the 
protons are charged. 
Taking $p_{v\p}^i=0$ we have
\begin{eqnarray}
{ p^{v\n}_\theta \over m_\n} = v^{v\n}_\theta + \varepsilon_{\n} (v^{v\p}_\theta - v^{v\n}_\theta) & = & \frac{\kappa}{2 \pi } \ , \\
{ p^{v\p}_\theta \over m_\p} = v^{v\p}_\theta + \varepsilon_{\p} (v^{v\n}_\theta - v^{v\p}_\theta) & = & 0 \label{pp0} \ ,
\end{eqnarray}
where $\kappa=h/2m_\n$. 
For later convenience we have identified the momentum associated with the vortex flow $ p_{v\X}^i$ 
by an additional index $v$. This will later allow us to distinguish between a uniform flow past, and 
the rotation induced by, the vortex.  
If we solve for the individual flows induced by the vortex we find
\begin{eqnarray}
v^{v\n}_\theta& = & \left(\frac{\varepsilon_{\p} - 1}{\varepsilon_{\p} + \varepsilon_{\n} - 1} \right)
\frac{\kappa}{2 \pi } \ , \label{vydef}\\
v^{v\p}_\theta & = &  \left(\frac{\varepsilon_{\p}}{\varepsilon_{\p} + \varepsilon_{\n} - 1} \right)
\frac{\kappa}{2 \pi } \ .
\label{vxdef}\end{eqnarray}

\section{The Magnus Force}

Having discussed the nature of the quantised vorticity, we are ready to investigate the force acting 
on a vortex due to a uniform flow past it. To do this, we require the flow of 
momentum $\pi_\X^i = n_\X p_\X^i$. That is, we need
\begin{eqnarray}\label{eq:MomComp}
\pd{\pi^{\X}_{i}}{t} & = & \pd{}{t} ( n_{\X} p^\X_i ) =  n_{\X} \pd{p_i^\X }{t}  +  p_i^\X \pd{n_{\X}}{t} \ .
\end{eqnarray}
Combining  (\ref{eq:Cont}) and (\ref{eq:MomEqu}) we can show that 
\begin{eqnarray}
\pd{}{t}(\pi_{i}^{\n} + \pi_{i}^{\p}) & = & - \rho \nabla_{i} \Phi - n_{\n} \nabla_{i} {\mu_{\n} - n_{\p} \nabla_{i} \mu_{\p}}
+ \alpha \nabla_{i} w_{\p\n}^2  \nonumber - \nabla_{j} [ n_{\n}p_i^\n v_\n^j + n_{\p}p_i^\p v_\p^j ] \\
& \approx &  - \nabla_j \left\{ n_{\n}p_i^\n v_\n^j + n_{\p}p_i^\p v_\p^j + \delta_i^j \left[ 
\rho \Phi + n_{\n} {\mu_{\n} + n_{\p} \mu_{\p}}
- \alpha w_{\p\n}^2
\right] \right\} \nonumber \\
& = & - \nabla_{j} \Pi^{j}_{i} \ ,
\label{Pidef}\end{eqnarray}
where $\rho = \rho_\n + \rho_\p = m_\n (n_\n+n_\p)$ is the total mass density --- we take $m_\n=m_\p$ throughout this paper.
It has been assumed that we are working on a sufficiently small scale that we can take the number densities
$n_\X$ and the entrainment parameter $\alpha$ constant.

In order to quantify  the force acting on a vortex it is natural to work in a frame 
in which the vortex is at rest. In that frame we want the background flow to be stationary and irrotational. 
Then we want to ask what the effects of introducing a single vortex in this flow may be.
Translating (\ref{eq:MomEqu}) into a frame moving with velocity $v^\L_i$ we get
\begin{eqnarray}
\left(\pd{}{t} + {w}^{j}_{\X\L}\nabla_{j} \right)   
\left[ {w}_{i}^{\X\L} + \varepsilon_{\X} w_i^{\Y\X} \right] + \nabla_{i} (\Phi + 
\tilde{\mu}_{\X}) 
+ \varepsilon_{\X}  w_j^{\Y\X} \nabla_{i} w^{j}_{\X\L} = 0 \ , 
\end{eqnarray}
where $w_i^{\X\L} = v_i^\X-v_i^\L$. 
Imposing the condition of a stationary and irrotational flow, the latter of which 
leads to 
\be
w_{\X\L}^j \nabla_i \left[  {w}_{j}^{\X\L} + \varepsilon_{\X} w_j^{\Y\X} \right] =
w_{\X\L}^j \nabla_j \left[  {w}_{i}^{\X\L} + \varepsilon_{\X} w_i^{\Y\X} \right] \ ,
\ee
we have
\be
\nabla_i \left\{{ 1 \over 2} w_{\X\L}^2 + \Phi + \tilde{\mu}_\X + \varepsilon_\X w_{\X\L}^j w_j^{\X\Y} \right\} = 0 \ .
\ee
After integration, this yields 
\be
{ 1 \over 2} w_{\X\L}^2 + \Phi + \tilde{\mu}_\X + \varepsilon_\X w_{\X\L}^j w_j^{\X\Y}  = C_\X = \mbox{ constant} \ .
\ee
From these two equations, we have
\be
D = \rho_\n C_\n + \rho_\p C_\p  =  { 1 \over 2} \rho_\n w_{\n\L}^2 +  
{ 1 \over 2} \rho_\p w_{\p\L}^2 + \rho\Phi + n_\n\mu_\n + n_\p\mu_\p
- 2\alpha w_{\p\n}^2 \ .
\ee
Combining this with (\ref{Pidef}), obviously translated into the vortex frame, we can show that
\begin{eqnarray}
\Pi_i^j &=& \rho_\n w_{\n\L}^j w_i^{\n\L} 
+ \rho_\p w_{\p\L}^j w_i^{\p\L}  - 2\alpha w_{\p\n}^j w_i^{\p\n}  + \delta_i^j \left[
D - { 1 \over 2} \rho_\n w_{\n\L}^2 -  { 1 \over 2} \rho_\p w_{\p\L}^2 + \alpha w_{\p\n}^2
 \right] \ .
\label{Pi2}\end{eqnarray}

The force per unit length that acts on the vortex
is now determined by the flow of momentum 
through a cylinder enclosing the vortex. That is, we need to evaluate
\begin{equation}\label{eq:ForceInteg}
f_{i} = \oint_{C} \Pi_{i}^{j} s_{j} dl \ ,
\end{equation}
where $C$ encloses the vortex and $s_j$ is the unit normal to the cylinder. 
For a single vortex we can assume the flow to be approximately of the form
\begin{equation}
 w_i^{\X\L} = U^{\X}_i + v^{v\X}_i \ ,
\end{equation}
where $U_i^\X$ is uniform, stationary and irrotational in the vortex frame.
The vortex flows $v_i^{v\X}$ are given by (\ref{vydef}) and (\ref{vxdef}).
Substituting in (\ref{Pi2}) and retaining only those terms that will eventually contribute to
the integral in  (\ref{eq:ForceInteg}) we find
\be
\Pi_i^j s_j = n_\n U_\n^j ( p_i^{v\n} s_j -  p_j^{v\n} s_i )  + n_\p U_\p^j ( p_i^{v\p} s_j  -  p_j^{v\p} s_i ) \ .
\ee
To arrive at this result we have used the fact that 
 all coefficients of $s_{i}$ which are constant on the contour
vanish when integrated around a circle.
As the vector $s_{i}$ in polar coordinates is proportional to the radial vector and $v_{v\X}^i$ is in the $\theta$ direction 
we obviously have $v_{v\X}^{j} s_{j} = 0$. Finally, 
as $U_\X^i$ is a constant flow, the integral around the vortex of $(U_\X^{j}s_{j})U^\X_{i}$ vanishes.
Our final expression can be rewritten as
\be
\Pi_i^j s_j = n_\n \epsilon_{ijk} U_\n^j ( \epsilon^{klm}p_l^{v\n} s_m ) \ ,
\ee
where we have used the fact that $ p_i^{v\p}=0$ for a single neutron vortex.

If we define the ``vorticity'' $\kappa^i$ as a vector with magnitude $\kappa$ which is aligned with 
$\epsilon^{ijk} \nabla_j p^\n_k$, i.e. use
\be
\epsilon_{ijk} p_{v\n}^j s^k = - {m_\n \kappa_i \over 2 \pi} \ ,
\ee
(in orthonormal cylindrical coordinates $\vec{\kappa} = \kappa \hat{e}_z$), and work out the force integral, 
we arrive at the final result for the  Magnus force acting on the vortex
\be
f_i^\mathrm{M} =  \rho_\n \epsilon_{ijk} U_\n^j \kappa^k =  -  \rho_\n \epsilon_{ijk} w_{\n\L}^j \kappa^k = \rho_\n \epsilon_{ijk} \kappa^j  w_{\n\L}^k \ . 
\label{magnus}\ee
From this result, 
the force density (per unit volume) on a collection of vortices follows readily as $n_v f_i^\mathrm{M}$. 

It is important to note that entrainment affects the Magnus force in two ways. 
First of all, it enters (\ref{magnus}) via $\kappa^j$. Secondly, it also impacts on the vortex number density 
$n_v$ according to (\ref{n_v}).
These results agree 
with the discussion of \citet{cc05}. Our final formula (\ref{magnus}) also agrees with the result used 
by 
\citet{lsc98}.  \citet{mendII} makes use of a more generic expression which allows for 
the presence of vortices in each different fluid. The coupling coefficients in that expression are, however, left
unspecified. 

\section{The Mutual Friction force}

Having found the form of the Magnus force, we can determine the ``mutual friction'' force, which represents a balance between the
Magnus force and standard ``resistivity'' due to electrons scattering off the magnetic field associated with each vortex. 
Taking the latter force to be proportional to the difference in velocity between the vortex and the charged fluid flow (protons and electrons), 
we have
\begin{equation}
f_{i}^\mathrm{e} = {\cal R} (v^{\p}_{i} - v^{\L}_{i}) \ .
\label{resist}\end{equation}
Assuming that the vortex can be treated as massless (see  \citet{mendI} for a justification of this assumption), this force must 
equal the Magnus force given by (\ref{magnus}). 
Solving the resultant equation for $v_i^\L$ (using repeated cross products with ${\kappa}^i$, see for example \citet{hv})
we find 
\begin{eqnarray}
v_{\L}^{i} & = & v_{\p}^{i} + \frac{{\cal R}}{\rho_{\n} \kappa^{2}} \left( \frac{1}{1 + {\cal R}^{2}/ \rho_{\n}^{2} \kappa^{2}} \right) \epsilon^{ijk} \kappa_{j}  w^{\p\n}_k  \nonumber \\
& & +  \frac{1}{\kappa^{2}} \left( \frac{1}{1 + {\cal R}^{2}/ \rho_{\n}^{2} \kappa^{2}} \right) \epsilon^{ijk} \kappa_{j} \epsilon_{klm} \kappa^{l} w_{\p\n}^m \ .
\label{mfforce1}\end{eqnarray}
Consequently, the force per unit length acting on the vortex is
\begin{eqnarray}
f^{i}_\mathrm{e} & = & {\cal R}( v_{\p}^{i} - v_{\L}^{i}) \nonumber \\
& = & \frac{{\cal R}^{2}}{\rho_{\n} \kappa^{2}} \left( \frac{1}{1 + {\cal R}^{2}/ \rho_{\n}^{2} \kappa^{2}} \right) \epsilon^{ijk} \kappa_{j}w^{\p\n}_k \nonumber \\
& & +  \frac{\cal R}{\kappa^{2}} \left( \frac{1}{1 + {\cal R}^{2}/ \rho_{\n}^{2} \kappa^{2}} \right) \epsilon^{ijk} \kappa_{j} \epsilon_{klm} \kappa^{l}w_{\p\n}^m \ .
\label{mfforce}\end{eqnarray}
The first term in this expression is analogous to the Magnus force, although now expressed in terms 
of the velocity difference $w_i^{\p\n}$. As this force is perpendicular to the relative velocity, it is non-dissipative. The second term, on the 
other hand, can be rewritten using
\be
\epsilon^{ijk} \kappa_{j} \epsilon_{klm} \kappa^{l}w_{\p\n}^m = \kappa^2 (\hat{\kappa}^i\hat{\kappa}^j - g^{ij}) w^{\p\n}_j  \ ,
\ee
where the bracket can be recognised as the projection orthogonal to $\hat{\kappa}^i= \kappa^i/\kappa$. This term induces dissipation in the flow.

From (\ref{mfforce1}) and (\ref{mfforce}) we also see that:
\begin{itemize}
\item In the limit ${\cal R} \to \infty$ we must have $v_i^\mathrm{\L} \to v_i^\p$. 
That is, the neutron vortices are strongly coupled to the charged fluid.

\item In the opposite, ``weak coupling'', limit where ${\cal R}\to 0$, the vortices must flow 
with the neutron fluid, i.e.,  we have $v_i^\mathrm{\L} = v_i^\n$.

\item The dissipative part of the mutual friction force, somewhat counterintuitively, 
vanishes in both of these limits. 

\end{itemize}

The form of the mutual friction to be used in the equations of motion is obtained, assuming that there is no direct 
vortex interaction (see \citet{ruder} for a discussion of such interactions), by multiplying (\ref{mfforce}) by the vortex density $n_v$. 
This is then the force that acts on the neutron superfluid, eg. which enters the right-hand side of Eq. (\ref{eq:MomEqu})
with $\X=\n$. An equal and opposite force acts on the charged
conglomerate, and provides the right-hand side of Eq. (\ref{eq:MomEqu}) with $\X=\p$. 

\section{Estimating the coefficients}

To complete our investigation, and in order to facilitate the use of our results in studies of
the dynamics of neutron stars, we need to discuss the parameters that
determine the strength of the mutual friction force. In essence, we need to estimate the 
``friction coefficient'' $\cal R$. To do this, we rely on the previous work of
\citet{als84} and  \citet{mendII}. Below we ``translate'' their analysis 
into our formalism. 

The contribution to the mutual friction force which is expected to provide the dominant coupling mechanism 
between the superfluid neutrons and the charged conglomerate is due to electrons scattering off the magnetic 
field associated with each vortex. An early analysis of this coupling was carried out by 
\citet{sss}, who discussed the importance of the spontaneous magnetisation of the vortex. 
Shortly after this analysis it was realised that the proton current induced by the entrainment effect 
would lead to a significantly stronger magnetic field  \citep{als84}. Hence, we focus our attention on this
case. 

For superconducting protons, the expression for the momentum must be replaced by (see, for example, \citet{prix2})
\be
p_i^\p  + { e \over c} A_i= { \hbar \over 2} \nabla_i \chi_\p \ .
\label{pmomA}\ee
(Here and in the following we are using Gaussian units). 
The magnetic field 
associated with the flow follows from, firstly the definition of the magnetic potential
\be
B_i = \epsilon_{ijk}\nabla^j A^k
\label{Bdef}\ee
and secondly, the Maxwell equation
\be
j_i = { c \over 4\pi} \epsilon_{ijk} \nabla^j B^k = { e n_\p v_i^\p} \ , 
\label{maxwell}
\ee
where the right-hand side is the charge current. In writing down this relation we have adopted the convention that the
charge currents affect only the magnetic induction $B_i$, see \citet{tilley} for further discussion. 
It is now straightforward to combine 
(\ref{Bdef}) and (\ref{maxwell}) to obtain the London equation for $A_i$. 

Presently, we are primarily interested in the magnetic field generated by the flow of entrained protons around a single neutron vortex. 
This means that it is natural to assume that there are no fluxtubes in the proton fluid. This is equivalent to assuming that 
the phase $\chi_\p$ is smooth. Then  a gauge transformation can be made such that (\ref{pmomA}) is replaced by, see \citet{tilley},
\be
p_i^\p  + { e \over c} A_i= 0 \ .
\label{pmom0}\ee
Thus it follows that 
\be
m_\p(1-\varepsilon_\p) v_i^\p = - { e \over c} A_i - m_\p \varepsilon_\p v_i^\n \ ,
\ee
and we arrive at the following equation for the 
magnetic flux (using the fact that
$\nabla_i B^i=0$ and assuming that the vortex is represented 
by a delta-function)
\be
\nabla^2 B_i - { 1 \over \Lambda_*^2} B_i = { 8 \pi e \alpha \over m_\p m_n c } 
\left( 1 - { 2 \alpha \rho \over \rho_\n \rho_\p} \right)^{-1} \epsilon_{ijk}\nabla^j p^k_\n \ .
\label{London}\ee
Here, the effective London penetration length $\Lambda_*$ is defined by
\be
\Lambda_*^2 = { c^2 m_\p^2 \over 4 \pi e^2 \rho_\p } \left( 1 - { 2 \alpha \rho \over \rho_\n \rho_\p} \right) 
\left( 1 - {2\alpha \over \rho_\n} \right)^{-1} \ .
\ee
Comparing (\ref{London}) to  equation (14) of \citet{als84}, and recalling the fact that the ``superfluid velocity''
in the orthodox formalism is in fact the rescaled momentum, we find 
that we should identify
\be
\rho_s^{\p\p} = \rho_\p \left( 1 - { 2 \alpha \rho \over \rho_\n \rho_\p} \right)^{-1} 
\left( 1 - {2\alpha \over \rho_\n} \right) \ , \qquad \mbox{ and } \qquad 
\rho_s^{\p\n} = - 2\alpha  \left( 1 - { 2 \alpha \rho \over \rho_\n \rho_\p} \right)^{-1} \ .
\ee
The relationship between the two formalisms has already been discuseed by  \citet{rotpaper}.
Using their analysis we readily demonstrate that the above identification is correct.

As the equations for the magnetic field are identical, the required solution is identical to 
that given in previous work \citep{als84,mendI}.  Solving (\ref{London}) for the case where the 
vortex is represented by a delta-function \citep{fetter}, we readily find that the only non-vanishing 
component of the magnetic field is 
\be
B^z = { \Phi_* \over 2 \pi \Lambda_*^2} K_0 (r/\Lambda_*) \ .
\ee
Here $K_0$ is a modified Bessel function,  
\be
\Phi_* = { hc \over 2e} { m_\p \over m_\n } { \rho_s^{\p\n} \over \rho_s^{\p\p}} \ ,
\ee
and $m_\p=m_\n$ for all practical purposes.
Crucially, this means that 
\be
|\vec{A}| \propto K_1(r/\Lambda_*) \sim \sqrt{\Lambda_* \over r} \exp(-r/\Lambda_*) \qquad \mbox{ for  } r>>\Lambda_* 
\ee
This resolves the apparent contradiction between (\ref{pp0}), which formed a key part of our derivation of the Magnus force, 
and the correct formula for charged protons, Eq.~(\ref{pmom0}). We see that the analysis in Section~III holds,
 provided that the force calculation 
can be performed sufficiently
far away from the vortex that $A_i$ can be neglected
yet close enough that the assumption of essentially constant densities, entrainment 
parameters etcetera holds. This should always be possible, since 
$\Lambda_*$ is many orders of magnitude smaller than eg. the intervortex separation.

For later convenience, it is useful to take a brief detour at this point and introduce 
the effective proton mass $m_\p^*$. This concept was also discussed by  \citet{rotpaper}. It is natural that the 
entrainment effect can be expressed in terms of an altered ``effective'' mass since it couples the momenta 
of the two fluids. The analysis leading to this is, in fact, very simple. 
In a frame comoving with the neutrons, i.e. in which $v_i^\n=0$,
(and well away from all vortices in a region where we can neglect the magnetic field, according to the discussion above) 
we have
\be
p_i^\p = m_\p (1- \varepsilon_\p) v_i^\p \equiv m_\p^* v_i^\p \ .
\ee 
Hence, it follows that
\be
2 \alpha = \rho_\p \varepsilon_\p =  n_\p (m_\p - m_\p^*) = n_\p \delta m_\p^* \ .
\ee
(Note that we could alternatively have defined the effective mass in the frame where $p_i^\n=0$. As discussed by  \citet{rotpaper}, 
the result is the same provided that the proton fraction is small --- the limit which is relevant for neutron star cores.). 

This means that we can use 
\be
\rho_s^{\n\p} = - \rho_\p \left( {\delta m_\p^* \over m_\p} \right) \left(  1 - { \rho \over \rho_\n}  {\delta m_\p^* \over m_\p} \right)^{-1} \approx
 - \rho_\p \left( {\delta m_\p^* \over m_\p^*} \right) \ ,
\ee
which should be accurate in the case of neutron stars where the proton fraction $x_\p=\rho_\p/\rho$ is small.
We also have
\be
\rho_s^{\p\p} = \rho_\p { \rho_\n m_\p - \rho_\p \delta m_\p^* \over \rho_\n m_\p - \rho \delta m_\p^* }  \approx \rho_\p \left( {m_\p \over m_\p^*} \right) \ .
\ee
These expressions are, not surprisingly, identical to those given by \citet{als84}.
Given these approximations, the penetration length is
\be
\Lambda_* \approx 1.3\times 10^2 \left[ \left( { x_\p \over 0.05}\right) \left({ \rho \over 10^{14} \mathrm{g/cm}^3 }\right)  { m_\p \over m_p^*} \right]^{-1/2} \ \mathrm{fm} \ ,
\ee 
and the magnetic field associated with the vortex core is approximated by \citep{als84}
\be
B \approx { |\Phi_*| \over 2 \pi \Lambda_*^2} \approx 1.9\times10^{14} \mathrm{G} \ \left( {x_\p \over 0.05}\right) \left({\rho \over 10^{14} \mathrm{g/cm}^3 } \right) \left| { \delta m_p^* \over m_p^*} \right| \ .
\ee

In order to estimate the relaxation time for electrons scattered off the vortex magnetic fields, we
combine three further results from \citet{als84}. The first is the relaxation timescale $\tau_0$ in the limit of a vanishing vortex radius.
It follows as
\be
\tau_0^{-1} = \pi N_\tau \Phi_*^2 \ ,
\ee
where
\be
N_\tau = { 2 \pi \over \hbar} n_v \left( {e\hbar \over 2m_\e c } \right)^2
\left( { m_\e c^2 \over E_{F\e} } \right)^2 { E_{F\e} \over (\pi \hbar c)^2 } \ .
\ee
We will assume that the electrons are ultrarelativistic, i.e. use
\be
E_{F\e} = \hbar c k_{F\e} ,  \quad \mbox{ where } \quad k_{F\e} = (3\pi^2n_\e)^{1/3} = (3\pi^2n_\p)^{1/3} \ .
\ee
As the electrons and protons are expected to couple on a much shorter timescale 
we account for the increased inertia by using
\be
\tau_v = \left( { m_\p c^2 \over \hbar c k_{F\e} }  \right) \tau_0 \ .
\ee 
The final factor encodes the dependence on the finite size of the scattering centre. 
As discussed by \citet{als84}, this leads to 
\be
\tau_v \longrightarrow { 16 \over 3\pi} { \alpha \over \beta} \tau_v \ ,
\ee
where 
\be 
{\alpha \over \beta} = 2 k_{F\e} \Lambda_* \approx 120 \left({m_\p^* \over m_\p} \right)^{1/2} \left[ 
\left( {x_\p \over 0.05} \right) \left({\rho_\p \over 10^{14} \mathrm{g/cm}^3 }\right)
\right]^{-1/6} \ .
\ee

Having arrived at an estimate of the timescale on which the vortices relax to the motion of 
the charged components, we can make connection with our expression for the mutual friction 
force from the previous section. To do this we note that the relative velocity between vortices and charged 
components relaxes according to $\partial_t \Delta v_i = - \Delta v_i /\tau_v$, 
from which we can deduce that the average force  (per unit length) acting on a typical vortex is
\be
\langle f_i \rangle = { \rho_\p \over n_v \tau_v} (  v^\p_i - v^\mathrm{L}_i) \ .
\ee
Comparing this to (\ref{resist}) we see that
\be
{\cal R} = { \rho_\p \over n_v \tau_v} \ .
\ee
As discussed in the previous section, it is useful to establish whether 
we are in the regime of strong or weak coupling. The above analysis
leads to the estimate 
\be
\left( { { \cal R} \over \rho_\n \kappa} \right)^2 \approx 1.6\times10^{-7} \left( {\delta m_\p^* \over m_\p} \right)^4
\left( {m_\p \over m_\p^*} \right)  \left( { x_\p \over 0.05} \right)^{7/3} \left( { \rho \over 10^{14} \mathrm{g/cm}^3} \right)^{1/3} << 1 \ .
\ee 
Given that the  effective proton mass is such that
$m_\p^*/m_\p\approx 0.5-0.7$ we are firmly in the weak coupling regime. 

Finally, let us return to (\ref{mfforce}) --- the force per unit length acting on an individual vortex --- and
replace it with the relevant force acting on the superfluid neutrons after averaging over the vortices. This is 
a slightly subtle issue. First we need to appreciate that the vortices are already accounted for in the averaged equations, eg. (\ref{mom_2}).
In a sense this means that the Magnus force (\ref{magnus}) is already contained in this equation. Hence we need to add only the 
resistive part (\ref{resist}) to the description. The force that we require thus follows simply by multiplying (\ref{mfforce}) by the local surface density of 
vortices $n_v$. Provided that we are dealing with the weak coupling limit, the resultant force acting on the neutron fluid 
can be written 

\be
f_\mathrm{mf}^i = {\cal B} \rho_\n n_v \epsilon^{ijk} \hat{\kappa}_j \epsilon_{klm} \kappa^l w^m_{\p\n} + 
{\cal B}^\prime \rho_\n n_v \epsilon^{ijk} \kappa_j w_k^{\p\n}  \ .
\label{mf_final}\ee
An equal and opposite force acts on the proton fluid. 
Here
\be
{\cal B} = { { \cal R} \over \rho_n \kappa} \approx  
4 \times10^{-4} \left( {\delta m_\p^* \over m_\p} \right)^2
\left( {m_\p \over m_\p^*} \right)^{1/2} \left( { x_\p \over 0.05} \right)^{7/6}  \left( { \rho \over 10^{14} \mathrm{g/cm}^3} \right)^{1/6} \ ,
\ee
and
\be
{\cal B}^\prime = {\cal B}^2\ .
\ee
are dimensionless parameters. These results should be compared to the parameters used by  \citet{mendII}, and it is easy to confirm that 
the two results are in perfect agreement.

\section{Concluding remarks}

In this paper we have derived the form of the vortex-mediated mutual friction, which arises as electrons scatter 
dissipatively off of the magnetic fields associated with the entrained proton currents and each neutron vortex, within the 
superfluid formalism developed by, for example, \citet{rp1,monster}. In doing this we have made contact with 
previous work based on the orthodox Landau formulation \citep{als84,mendI}, and demonstrated that the 
two pictures are consistent. Since our description incorporates the entrainment effect in a 
transparent way (by making the appropriate distinction between true transport velocities and momenta)
this comparison lends strong support not only to our present results but to the previous work as well.

To conclude our discussion, let us put the final expression for the mutual friction force (\ref{mf_final}) to 
use by working out the timescale on which a difference in rotation between the two fluids 
in a neutron star core (eg. following a pulsar glitch)
is relaxed locally. 
Given (\ref{mom_2}) and (\ref{mf_final})  we see that the system evolves according to
\be
\left. \begin{array}{ll} 
\n_\n \partial_t p_i^\n + \ldots = f^\mathrm{mf}_i \\
n_\p \partial_t p_i^\p + \ldots = - f^\mathrm{mf}_i \end{array} \right\} \longrightarrow  {  m_\p^* \over m_\p} 
\partial_t w_i^{\n\p} + \ldots  \approx -{  { \cal B} \kappa n_v \over x_\p} w_i^{\n\p} \ ,
\ee
where, on the left-hand side  we have used the definition of  the momenta and assumed that
the proton fraction is small (in accordance with the preceding analysis), and on the right-hand side we implicity 
assume that the velocity difference is perpendicular to the vortices (that is, the two fluids rotate around the same axis). 
From this expression we see that the timescale on which the two fluids are dynamically coupled can be 
estimated by
\be
\tau_d \approx  {  m_\p^* \over m_\p} { x_\p \over {\cal B} \kappa n_v}  \ .
\ee
Taking $n_v \kappa \approx 4\pi /P$, i.e. assuming that the two rotation rates are similar
to the observed pulsar period $P$, cf. (\ref{n_v}), we have
\be
\tau_d \approx 10 P (s)
\left( { m_\p^* \over \delta m_p^*} \right)^2  \left( {x_\p \over 0.05} \right)^{-1/6} \left({ \rho \over 10^{14} \mathrm{g/cm}^3} 
\right)^{-1/6} \ .
\ee
This estimate is about
one order of magnitude smaller than the classic result of \citet{as88}. That the two results differ is perhaps not too 
 surprising. After all, the \citet{as88}  analysis was based on 
an explicit solution for the motion of an individual vortex affected by the Magnus force (\ref{magnus}) and the resistivity (\ref{resist})
given a constant relative rotation rate.
In contrast, our estimate does not refer to the explicit vortex motion, only to the way that the two fluids couple
via the vortices.  Of course, the main astrophysical conclusion remains unaltered.
The coupling timescale is much shorter than the observed relaxation timescale following (say) 
the large Vela glitches \citep{as88}.
This suggests that the glitches are unlikely to be associated with 
the core and points instead to the superfluid neutrons in the crust 
playing a key role.

Basically, we have now prepared the ground for discussions of the relevance of mutual friction 
in different astrophysical scenarios 
within our formalism. This is a very important step forwards since it allows us to consider 
key problems concerning, for example, the mutual friction damping of pulsation modes driven unstable 
by gravitational radiation. As the answer may be of significance for gravitational-wave 
observations, the available results in that problem area \citep{lm_f,lm_r} must be verified 
by independent work. Given the present analysis, we are set to carry out such  calculations and expect to report the results 
in the not too distant future. 
 
\section*{Acknowledgements}

NA acknowledges support from PPARC via grant no PPA/G/S/2002/00038 and Senior Research Fellowship no
PP/C505791/1. GLC is supported by NSF grant no PHY-0457072.

%\bibitem
%[\protect\citepauthoryear{{}}{{}}{}]
%{narev}
%{Andersson} N.,  2003, Class. Quantum Grav., 20, R105

\end{document}